\documentclass{vldb}
\usepackage{graphicx}
\usepackage{amsmath}
\usepackage{epsfig}
\usepackage{url}
\usepackage{multirow}
\usepackage{color}
\usepackage{ulem}
\usepackage{balance}
\usepackage{subfigure}
\usepackage{varioref}
\usepackage{bold-extra} 
\usepackage{xspace}
\usepackage[colorinlistoftodos]{todonotes}
\usepackage{needspace}
\reversemarginpar

\usepackage{algorithmic} 
\usepackage{algorithm}

\newcommand{\map}{\textsc{Map}\xspace}

\newcommand{\reduce}{\textsc{Reduce}\xspace}

\newcommand{\shuffle}{\textsc{Shuffle}\xspace}

\newcommand{\hjt}{\textsc{JobTracker}\xspace}
\newcommand{\htt}{\textsc{TaskTracker}\xspace}
\newcommand{\sus}{\textsc{Suspend}\xspace}
\newcommand{\res}{\textsc{Resume}\xspace}
\newcommand{\kil}{\textsc{Kill}\xspace}
\newcommand{\wait}{\textsc{Wait}\xspace}

\newcommand{\parag}[1]{\paragraph*{#1}}

\newcommand{\SKIP}[1]{}

\definecolor{OliveGreen}{cmyk}{0.64,0,0.95,0.40}

\begin{document}

%





\title{Practical Size-based Scheduling for MapReduce Workloads}

\numberofauthors{5}

\author{
\alignauthor
Mario Pastorelli\\
       \affaddr{EURECOM}\\
       \affaddr{Campus SophiaTech, France}\\
       \email{pastorel@eurecom.fr}
\alignauthor
Antonio Barbuzzi\\
       \affaddr{EURECOM}\\
       \affaddr{Campus SophiaTech, France}\\
       \email{barbuzzi@eurecom.fr}
\and
\alignauthor Damiano Carra\\
       \affaddr{University of Verona}\\
       \affaddr{Italy}\\
       \email{damiano.carra@univr.it}
\alignauthor Matteo Dell'Amico\\
       \affaddr{EURECOM}\\
       \affaddr{Campus SophiaTech, France}\\
       \email{dellamic@eurecom.fr}
\alignauthor Pietro Michiardi\\
       \affaddr{EURECOM}\\
       \affaddr{Campus SophiaTech, France}\\
       \email{michiard@eurecom.fr}
}

\maketitle

\begin{abstract}
  We present the Hadoop Fair Sojourn Protocol
    (HFSP) scheduler, which implements a size-based scheduling
  discipline for Hadoop. The benefits of size-based scheduling
  disciplines are well recognized in a variety of contexts (computer
  networks, operating systems, etc...), yet, their practical implementation
  for a system such as Hadoop raises a number of important
  challenges.

  With HFSP, which is available as an open-source project, we address issues related to job size estimation, resource management
  and study the effects of a variety of preemption
  strategies. Although the architecture underlying HFSP is suitable
  for any size-based scheduling discipline, in this work we revisit
  and extend the Fair Sojourn Protocol, which solves problems
  related to job starvation that affect FIFO, Processor Sharing and a
  range of size-based disciplines.

  Our experiments, in which we compare HFSP to standard Hadoop schedulers, pinpoint at a significant decrease in average job sojourn times -- a metric that accounts for the total time a job spends in the system, including waiting and serving times -- for realistic workloads that we generate according to production traces available in literature.
\end{abstract}



\section{Introduction}
\label{sec:introduction}

The advent of large-scale data analytics, fostered by parallel processing frameworks such as MapReduce \cite{mapreduce}, has created the need to organize and manage the resources of clusters of computers that operate in a shared, multi-tenant environment. For example, within the same company, many users \textit{share} the same cluster because this avoids redundancy (both in physical deployments and in data storage) and may represent enormous cost savings. Initially designed for few and very large batch processing jobs, data-intensive scalable computing frameworks such as MapReduce are nowadays used by many companies for production, recurrent and even experimental data analysis jobs. This is substantiated by recent studies \cite{workloads, workloads_research} that analyze a variety of production-level workloads (both in the industry and in academia): an important characteristic that emerges from such works is that there exists a stringent need for \textit{interactivity}. The number of small jobs might be dominant in current workloads: these are preliminary data analysis tasks involving a human in the loop, which for example seeks at tuning algorithm parameters with a trial-and-error process, or even small jobs that are part of orchestration frameworks whose goal is to launch other jobs according to a workflow schedule.

In this work, we study the problem of job \textit{scheduling}, that is how to allocate the resources of a cluster to a number of concurrent jobs submitted by the users, and focus on the open-source implementation of MapReduce, namely Hadoop \cite{hadoop}. In addition to the default, first-in-first-out (FIFO) scheduler implemented in Hadoop, recently, several alternatives \cite{eurosys10, infocom11, nsdi11, deadlines, dynamic, flex} have been proposed to enhance scheduling: in general, existing approaches aim at two key objectives, namely \textit{fairness} among jobs and \textit{performance} in terms of job execution time. 

We propose to use job \textit{sojourn time} as a performance metric,
which accounts for the time a job spends in the system waiting to be
served and its execution time. We thus proceed with the design and implementation of a new scheduling protocol that caters both to a fair and efficient utilization of cluster resources, while striving to achieve short sojourn times. As such, our approach satisfies \textit{both} the interactivity requirements of small jobs, and the performance requirements of large batch jobs.

Our solution, Hadoop Fair Sojourn Protocol (HFSP), belongs to the
category of size-based, preemptive scheduling disciplines.  In
addition to addressing the problem of scheduling jobs characterized by
a complex structure in a multi-processor system, we propose an
efficient method to implement size-based scheduling when job size
cannot be known \textit{a priori}.

HFSP allocates cluster resources such that job size information is inferred while the job makes progress toward its completion. The scheduling discipline benefits from preemption to achieve short job sojourn times; however, preemption is not currently implemented in Hadoop. As such, we introduce a new set of primitives that enables HFSP to interrupt and eventually resume running jobs, and show in which cases this approach is superior to the widely adopted technique of killing running tasks to make room for other jobs. The contribution of our work can be summarized as follows:
\begin{itemize}
\item We design and implement the system architecture of HFSP, including a (pluggable) component to estimate job sizes, a dynamic resource allocation mechanism that strives at efficient cluster utilization and a new set of low-level primitives that allows preemptive disciplines. HFSP is available as an open-source project.\footnote{\url{https://bitbucket.org/bigfootproject/hfsp-project}}

\item HFSP uses a new scheduling discipline inspired by the Fair Sojourn Protocol \cite{fsp}. Job scheduling operates in a multi-processor context and aims to minimize job sojourn times. One of the main consequences of the HFSP discipline is that small jobs, for which ``interactivity'' is important, do not wait for a long time before being awarded cluster resources. The HFSP scheduler is also beneficial to medium-large jobs which are granted a large fraction of cluster resources.

\item We perform an extensive experiment campaign, where we compare the HFSP scheduler to the two main schedulers used in production-level Hadoop deployments, namely the FIFO and the FAIR schedulers. For the experiments, we use state-of-the-art workload suite generators that take as input realistic workload traces. Our results -- that we obtain on a large cluster deployed in Amazon EC2 -- show that the average job sojourn time achieved with HFSP is drastically reduced with respect to the other scheduler we examined. In addition, we show results that substantiate the claim of an efficient cluster resource utilization under heavy loads.
\end{itemize}

The remainder of the paper is organized as follows: in Section~\ref{sec:background} we provide background information on a set of scheduling disciplines and on some details of Hadoop MapReduce. In Section~\ref{sec:system} we describe in details the HFSP schedulers and its inner components.  We evaluate the performance of our job scheduler in Section~\ref{sec:experiments} and provide in Section~\ref{sec:discussion} additional considerations. In Section~\ref{sec:related_work} we discuss the related work, and we conclude in Section\,\ref{sec:conclusion}.

\section{Background}
\label{sec:background}
When comparing different scheduling disciplines, there are different performance metrics one can consider. In this work we focus on the \textit{mean response time} -- \textit{i.e.} the total time spent in the system, given by the waiting and service time, called also \textit{sojourn time} -- for each job, and \textit{fairness}. Next, we consider two disciplines that are relevant in our context: one that minimizes the mean response time and one that provides perfect fairness.

The optimal preemptive scheduling policy that minimizes the mean response time is the Shortest Remaining Processing Time (SRPT), where the job in service is the one with the smallest remaining processing time -- this policy requires the job size to be known \textit{a priori}. SRPT provides no guarantees on system fairness: as such, long jobs may starve. As opposed to minimizing the mean response time, the Processor Sharing\footnote{In this work, we compare our proposal to the FAIR scheduler, which implements the concept of PS, albeit with additional features to cope with a multi-processor system.} (PS) discipline is conceived to guarantee a fair share of system resources to be dedicated to each job: if $N$ jobs need to be served, with PS each receives a $1/N$th fraction of the system resources. However, the mean response time achieved by PS is higher than that obtained with SRPT.

Friedman and Henderson~\cite{fsp} study a scheduling policy for a single-server queue model that strives to obtain both (near) optimal mean response times and fairness, called Fair Sojourn Protocol (FSP). Since our work is inspired by FSP, in the following we provide sufficient background to understand its properties. 

\subsection{How FSP Works}
\label{sec:background_fsp}

The main idea of FSP is to run jobs in series rather than concurrently. FSP computes the completion time for each job under a virtual PS discipline. The order at which jobs complete in PS is used as a reference to schedule jobs in series. In the basic single server-queue model, this means that at most one job is served at a time, and that such job may be preempted by a newly arrived job. An example is the best way to illustrate how FSP works.

Assume that there are three jobs, $j_1$, $j_2$ and $j_3$, each requiring \textit{all the resources} available in the system. Such jobs arrive at time $t_1 = 0$s, $t_2 = 10$s and $t_3 = 15$s respectively; it takes 30 seconds to process job $j_1$, 10 seconds to process job $j_2$ and 10 seconds to process job $j_3$ (if all the resources are used, otherwise the time increases inversely proportionally to the available resources).

\begin{figure}[htbp]
\centering
       \includegraphics[width=0.98\linewidth]{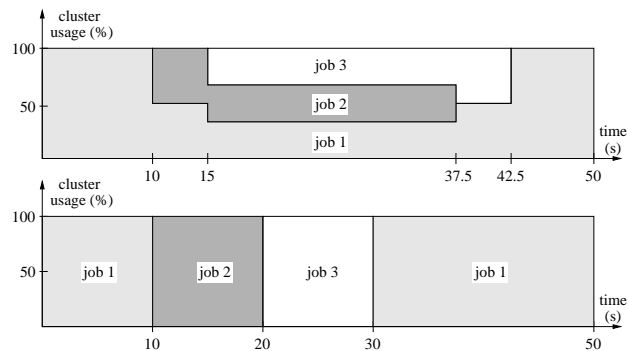}
\caption{Comparison between PS (top) and FSP (bottom).}
\label{fig:ps_vs_fsp}
\end{figure}

Figure\,\ref{fig:ps_vs_fsp} (top) represents the system utilization over time under the PS discipline: when job $j_2$ arrives, the server is shared between $j_1$ and $j_2$, and, when job $j_3$ arrives, the server is shared among the three jobs. The job completion order is $j_2$, $j_3$ and $j_1$. The bottom part of the figure shows how the workload described above is scheduled under the FSP discipline. When job $j_2$ arrives, since it would finish before job $j_1$ in case of PS, it preempts job $j_1$. When job $j_3$ arrives, it does not preempt job $j_2$, since it would finish after it in case of PS; when job $j_2$ finishes, job $j_3$ is scheduled since it would finish before job $j_1$ in case of PS. The FSP discipline ensures each job receives a fair amount of system resources, as when PS scheduling is used. At the same time, under FSP, the mean job completion time is considerably smaller than under PS. Next, using a simple example, we anticipate a more elaborate setup that underlies our work, whereas in Section~\ref{sec:scheduling} we detail all the hidden intricacies of a multi-processor version of FSP.

Assume that jobs $j_1$, $j_2$ and $j_3$ require 100\%, 55\% and 35\% of the system resources respectively. The arrival times are $t_1 = 0$s, $t_2 = 10$s and $t_3 = 13$s and the processing time (if the required share of system resources is given to each job) is 30 seconds for job $j_1$, 10 seconds for job $j_2$ and 10 seconds for job $j_3$.

\begin{figure}[htbp]
\centering
       \includegraphics[width=0.98\linewidth]{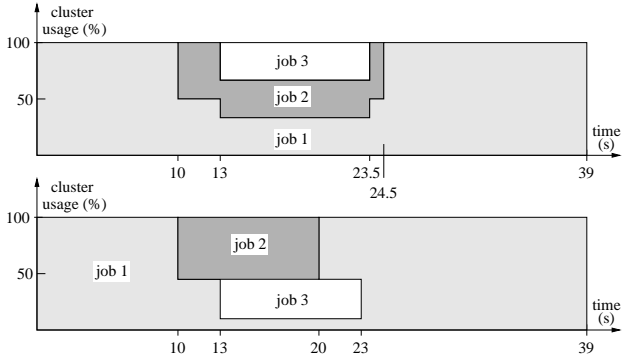}
       \caption{Comparison between PS (top) and an ideal multi-processor FSP (bottom), with
         jobs that do not require the full cluster.}
\label{fig:ps_vs_fsp_2}
\end{figure}

Figure\,\ref{fig:ps_vs_fsp_2} compares PS (top) to an ideal, \textit{multi-processor} version of FSP (bottom). With ideal FSP, job $j_2$ would preempt job $j_1$; since $j_2$ requires only 55\% of the system, the remaining 45\% can still be used by $j_1$. When job $j_3$ arrives, it would preempt job $j_1$ (but not job $j_2$), but it is sufficient to allocate 35\% of the system to serve it, leaving 10\% of the server to job $j_1$. As shown in the Figure, the mean job completion time under multi-processor FSP is smaller than that achieved by PS, and system resources are allocated such that no job is ``mistreated.''

 
\subsection{Hadoop MapReduce}
\label{sec:background_mr}

MapReduce, popularized by Google with their work in \cite{mapreduce}
and by Hadoop \cite{hadoop}, is both a programming model and an
execution framework. In MapReduce, a job consists of three phases and
accepts as input a dataset, appropriately partitioned and stored in a
distributed file system. In the first phase, called \map{}, a user-defined function is applied in
parallel to input partitions to produce intermediate data stored on
the local file system of each machine of the cluster; intermediate
data is sorted and partitioned when written to
disk. Afterwards, a \reduce phase is scheduled. It
  comprises a \shuffle sub-phase, where intermediate data is
  ``routed'' to the \textit{reducers}. Subsequently, intermediate data
  from multiple mappers is sorted and aggregated to produce output
  data which is written back on the distributed file system.

\parag{Hadoop Scheduling} In Hadoop, the \hjt{} takes care of
coordinating \htt{} nodes, which can be thought of as worker
machines. The scheduler, which is the subject of this work, resides in
the \hjt.  The role of the scheduler in MapReduce is to allocate
\htt{} resources to running tasks: \map{} and \reduce{} tasks are
granted independent \textit{slots} on each machine. The number of
\map{} and \reduce{} slots on each \htt{} is a configurable parameter,
which depends on the cluster in which Hadoop is deployed, and on the
characteristics (e.g., the number of CPU cores) of each server in the
cluster.

When a \textit{single} job is submitted to the cluster, the scheduler simply assigns as many \map{} tasks as the number of available slots in the cluster. Note that the total number of \map{} tasks is equal to the number of partitions of the input data. The scheduler tries to assign \map{} tasks to slots available on machines in which the underlying storage layer holds the input intended to be processed, a concept called \textit{data locality}. Also, the scheduler may need to wait for a portion of \map{} tasks to finish before scheduling subsequent mappers, that is, the \map{} phase may execute in multiple ``waves'', especially when processing very large data. Similarly, \reduce{} tasks are scheduled once intermediate data, output from mappers, is available.\footnote{Precisely, a configuration parameter indicates the fraction of mappers that are required to finish before reducers are awarded an execution slot.} When \textit{multiple} jobs are submitted to the cluster, the scheduler decides how to allocate available task slots across jobs.

The default scheduler in Hadoop implements a FIFO policy: the whole cluster is dedicated to individual jobs in sequence; optionally, it is possible to define priorities associated to jobs. In practice, the FIFO scheduler works as follows: it assigns tasks (\map{} or \reduce{}) in response to heartbeats sent by each individual \htt{}, which reports the number of free \map{} and \reduce{} slots available for new tasks. Task assignment is accomplished by scanning through all jobs that are waiting to be scheduled, in order of priority and job submission time. The goal is to find a job with a pending task of the required type (\map{} or \reduce{}). In particular, for \map{} tasks, once the scheduler chooses a job, it will select greedily the more suitable task to achieve \textit{data locality}. In this work we also consider the Hadoop Fair Scheduler, which we call FAIR. FAIR groups jobs into ``pools'' (generally corresponding to users or groups of users)and assigns each pool a guaranteed minimum share of cluster resources, which are split up among the jobs in each pool. In case of excess capacity (because the cluster is over dimensioned with respect to its workload, or because the workload is lightweight), FAIR splits it evenly between jobs. When a slot on a machine is free and needs to be assigned a task, FAIR proceeds as follows: if there is any job below its minimum share, it schedules a task of that particular job. Otherwise, FAIR schedules a task that belongs to the job that has received less resources, based on the notion of ``deficit.''

\section{Hadoop Fair Sojourn Protocol} \label{sec:system}

HFSP is a concrete implementation of the multi-processor FSP
introduced in Section~\ref{sec:background_fsp}. The abstract idea is
simple: prioritize jobs according to the completion time they would
have using processor sharing, and always use preemption to allocate
resources to the highest-priority job. The size of new jobs is at
first estimated roughly based on their number of tasks; by observing
the running time of the first few tasks, that estimate is then
updated. Implementing HFSP in practice, though, raises a variety of
issues; in the following we detail how we tackled them.

\subsection{The Job Scheduler}\label{sec:scheduling}

HFSP runs on a multi-processor system, managing jobs that may require
a different number of \map{} and \reduce{} tasks: unlike FSP and like
the multi-processor variant illustrated in
Figure~\vref{fig:ps_vs_fsp_2}, a job might need only a fraction of the
resources of a cluster, and therefore two or more jobs may be
scheduled at the same time.

\parag{Job Dependencies} In MapReduce, a job is composed by a \map{}
phase followed optionally by a \reduce{} phase. We estimate job size
by observing the time needed to compute the first few ``training''
tasks of each phase; for this reason we cannot estimate the length of
the \reduce{} phase when scheduling \map{} tasks. Because of this, for
the purpose of scheduling choices we consider \map{} and \reduce{}
phases as two separate jobs. For ease of exposition, we
  will thus refer to both \map and \reduce phases as ``jobs'' in the
  remainder of Section~\ref{sec:system}. As we experimentally show in
Section~\ref{sec:experiments}, the good properties of size-based
scheduling ensure shorter mean response time for both the \map{} and
the \reduce{} phase, resulting of course in better response times
overall.

\parag{Scheduling Policy} HFSP works by
estimating the size of each job using the training module described
in Section~\ref{sec:training}; this size is then used to compute the
completion time of each job in a simulated processor-sharing scheduler
(see Section~\ref{sec:virtual_cluster}).  When jobs arrive, all cluster resources are assigned to fulfill demands of jobs ranked by increasing simulated completion time. 
If a job ends up having more running tasks than assigned, excess slots
are released by using the preemption mechanisms described in
Section~\ref{sec:preemption}.

\parag{Training Phase} Initially, the training module
provides a rough estimate for the size of new jobs
(Section~\ref{sec:initial-estimation}). This estimate is then updated
after the first $s$ ``sample'' tasks of a job are executed
(Sections~\ref{sec:estimation-map} and~\ref{sec:estimation-reduce}).
To guarantee that job size estimates quickly converge to more accurate
values, the first $s$ tasks of each jobs are prioritized, and the
above mechanism is amended so that a configurable number of \map{} and
\reduce{} ``training'' slots are always available, if needed, for the
sample tasks. This number is limited in order to avoid starvation for
``regular'' jobs in case of a bursty arrival of a large number of
jobs.

\parag{Data locality} In order to minimize data transfer
latencies and network congestion, it is important to make sure that
\map{} tasks work on data available locally. For this reason, HFSP
uses the \textit{delay scheduling} strategy~\cite{eurosys10}, which
postpones scheduling tasks operating on non-local data for a fixed
amount of attempts; in those cases, tasks of jobs with lower priority
are scheduled instead.

\subsection{Job Size Estimation} \label{sec:training}

Size-based scheduling strategies require knowing the size of jobs. In
Hadoop, we do not have a perfect knowledge of the size of a job until
it is finished; however, we can at first estimate job size based on
characteristics known \textit{a priori} such as the number of tasks;
after the first sample tasks have executed, the estimation can be
updated based on their running time.

This estimation component has been designed to result in minimized
response time rather than coming up with perfectly accurate estimates
of job length; this is the reason why sample jobs should not be too
many (our default is $s=5$), and they are scheduled quickly. We stress
that the computation performed by the sample tasks is \textit{not}
thrown away: the results computed by sample tasks will be used in the
rest of the job exactly as the other regular tasks.

\subsubsection{Initial Estimation}
\label{sec:initial-estimation}

When a new job arrives in the system, a first rough estimate is used;
after obtaining enough information from the sample tasks, this
estimate will be updated as described in the following sections. In
Hadoop, the number of \map{} and \reduce{} tasks each job needs is
known \textit{a priori}. In turn, each \map{} task processes data
residing on a single, fixed-size, HDFS block. Our first approximation
will be therefore directly proportional to the number of tasks per
job.

The size of a \map{} (resp.~\reduce{}) job with $k$ tasks is, at
first, estimated as $\xi \cdot k \cdot l$, where $l$ is the average
size of past \map{} (resp.~\reduce) tasks, and $\xi \in [1, \infty]$
is a tunable parameter that represents the propensity that the system
has to schedule jobs of unknown size. At the extreme $\xi=1$, new jobs
are scheduled quite aggressively based on the initial estimate, with the possible drawback of scheduling particularly
large jobs too early.  At the extreme $\xi=\infty$, conversely,
non-sample tasks are only scheduled if all jobs which are not in the
training phase are satisfied. This avoids resorting to preemption for
jobs that turn out to be larger than at first envisioned, but this
choice may penalize small jobs that could be scheduled quickly as they
arrive in the scheduler, increasing the response time by adding
so-called ``waves'' (i.e., tasks scheduled after a first set of tasks
is already completed).

We note that these problems do not generally apply to particularly
small jobs having $s$ or less tasks, since all tasks are tagged as
sample and in most cases scheduled quickly.

\subsubsection{\map{} Phase Size}
\label{sec:estimation-map}

It has been observed \cite{eurosys10,swim}, across a variety of jobs,
that \map{} task execution times are generally stable and short, with
execution times around one minute. For this reason, it is possible to
perform job size estimation when the $s$ sample tasks are
completed.  Our estimation builds an approximate cumulative
distribution function (CDF) with the data points obtained by
measuring, for each sample task $j$ of job $i$, the execution time
$\sigma_{i,j}$; via regression analysis (in particular, least-square
fitting), that approximate CDF is then fitted to a target parametric
distribution. The estimated \map{} phase size is finally computed as
$k$ times the expected value of the resulting fitted distribution,
where $k$ is the number of \map{} tasks for the estimated job.

\parag{Data Locality} A \map sample task could perform
worse than normal due to network latencies if operating on non-local
data. However, since the sample tasks are between the first to be
scheduled, there is a larger choice of blocks to process, making the
need of operating on remote data less likely. In combination with the
delay scheduling strategy described in Section~\ref{sec:scheduling}, we
found that data locality issues on sample tasks, as a result, are
negligible.

\subsubsection{\reduce{} Phase Size}
\label{sec:estimation-reduce}

The \reduce{} phase can be broken down in two parts: \shuffle{} time
-- needed to move and merge data from mappers to reducers, and the
execution time of the \reduce{} function, which can only start when the \shuffle phase has completed.

\parag{Size of \shuffle} In general, as soon as a \reduce task is
scheduled, it starts pulling data from the output of mappers; once
data from all mappers is available, a \textit{global sort} is
performed by merging all the mappers' output. Since each mapper output
is already locally sorted, a simple $O(n)$ merge step is
sufficient. Network transfer and sort time are essentially
proportional to the size of the data to move, and the \shuffle{}
performance depends mostly on the amount of data and the
characteristics of the cluster it is run on, rather than on the type
of data that is moved.

For this reason, an approximate duration of the \shuffle phase can be
computed as follows. \textit{For each} of the $s$ sample \reduce tasks
of a job that enters the training module, we measure the time required
for their \shuffle phase. This is given by the difference between the
moment a task executes the \reduce{} function, and the moment the same
task was scheduled in the training module. The estimated \shuffle{}
time of the entire \reduce{} phase is then the \textit{weighted
  average} of the individual \shuffle times of the sample tasks
multiplied by the total number of \reduce tasks of the job, where the
weights are the normalized input data size to each sample task.

\parag{Execution Time} The execution time of the \reduce{}
phase is evaluated analogously to the \map phase described in
Section~\ref{sec:estimation-map}. However, \reduce tasks can be orders
of magnitude longer than \map tasks, therefore we aim at providing an
estimate of the duration of the sample tasks before their
completion. In particular, we set a timeout $\Delta$ (60s by
default). If a sample task $j$ of job $i$ is not completed by the
timeout, its estimated execution time will be
$\tilde{\sigma}_{i,j}=\frac{\Delta}{p_{i,j}}$, 
where $p_{i,j}$ is the \textit{progress} done during the execution
stage. The progress of a task is computed as the fraction of data
processed by a \reduce{} task over the total amount of its input
data.

Once we obtain the size (or an estimation of it) for each sample task,
we compute the total execution time using the same procedure described
in Section~\ref{sec:estimation-map}, using regression analysis to fit
measured data to a parametric distribution, and multiplying the
expected value of the resulting distribution by the number of \reduce
tasks in the job. The final estimate of the whole \reduce phase is
obtained by adding the estimated \shuffle time to this estimated
execution time.

\subsection{Virtual Cluster} \label{sec:virtual_cluster}

The estimated job size is expressed in a ``serialized form'', that is
the sum of runtimes of each of its tasks. This is useful because the
physical configuration of the cluster does not influence the estimated
size; moreover, in case of failures, the number of available \htt{}s
varies, but the size of jobs does not change.  The job scheduler,
though, requires an estimated completion time that depends on the
physical configuration of the real cluster. We obtain that by
simulating a processor sharing discipline applied on a \textit{virtual
  cluster} having the same number of machines and the same
configuration of slots per machine as the real cluster. The projected
finish time obtained in the virtual cluster is then fed to the job
scheduler, which will use it to perform its scheduling choices.



\parag{Resource Allocation} Virtual cluster resources need to be allocated following the principle of a fair queuing discipline. Since jobs may require less than their fair share, in HFSP, resource allocation in the virtual cluster uses a \textit{max-min fairness} discipline. Max-min fairness is achieved using a round-robin mechanism that starts allocating virtual cluster resources to small jobs (in terms of their number of tasks). As such, small jobs are implicitly given priority in the virtual cluster, which reinforces the idea of scheduling small jobs as soon as possible.

\parag{Job Aging} The HFSP algorithm keeps track of, in the virtual
cluster, the amount of work done by each job in the system. Each job
arrival or task completion triggers a call to the job aging
function, which uses the time difference between two consecutive
events as a basis to distribute progress among each job currently
scheduled in the virtual cluster.  In practice, each running task in
the virtual cluster makes a progress corresponding to the above time
interval. Hence, the ``serialized'' representation of the remaining
amount of work for the job is updated by subtracting the sum of the
progress of all its running tasks in the virtual cluster.

\subsection{Job Preemption} \label{sec:preemption} 
The HFSP scheduler uses preemption: a new job can suspend tasks of a
running job, which are then resumed when resources become
available. However, traditional preemption primitives are not readily
available in Hadoop. The most commonly used technique to implement
preemption in Hadoop is ``killing'' tasks
or entire jobs. Clearly, this is not optimal, because it wastes work,
including CPU and I/O. Alternatively, it is possible to \wait{} for a
running task to complete, as done by Zaharia \textit{et
  al.}~\cite{eurosys10}. If the runtime of the task is small, then the
waiting time is limited, which makes \wait{} appealing. In fact, we
suspend jobs using the \wait primitive for \map tasks which are
generally small. However, for tasks with long runtime, the delay
introduced by this approach may be too high.

\parag{Eager Preemption} Since \reduce tasks may be orders of
magnitude longer than \map tasks, to preempt \reduce jobs we adopt a
more traditional approach, which we name \textit{eager preemption}:
tasks are suspended in favor of other jobs, and resumed only when
their job is later awarded resources. Eager preemption requires
implementing \sus{} and \res{} primitives: we delegate them to the
operating system (OS).
\map and \reduce tasks are launched by the \htt as child Java Virtual
Machines (JVMs); child JVMs are effectively processes, which we suspend
and resume 
using standard POSIX signals: \texttt{SIGSTOP} and
\texttt{SIGCONT}. When HFSP suspends a task, the underlying OS
proceeds with its materialization on the secondary storage (in the
swap partition), \textit{if and when} its memory is needed by another
process. We note that our implementation introduces a new set of
states associated to an Hadoop task, plus the related messages for the
\hjt{} and \htt{} to communicate state changes and their
synchronization.

\parag{Task Suspension} As discussed in
Section~\ref{sec:scheduling}, the job scheduler allocates cluster
resources to jobs that finish first, as computed in the virtual
cluster. A new job arriving in the system may induce -- depending on
its size -- the job scheduler to \sus{} a running job. In practice,
the job scheduler suspends tasks, rather than jobs: task suspension
works as follows. Upon reception of a heart-beat from a \htt, the
job scheduler verifies whether a job tagged for suspension occupies
resources. If this is the case, it suspends enough tasks of that job
to provide the needed resources to the job that should take its place.

It may happen that not all the tasks of a job have to be suspended: in
that case, tasks that have been launched last are the ones which get
suspended, for two practical reasons: i) ``oldest'' tasks are the most
likely ones to finish first, freeing resources to other tasks; ii)
``young'' tasks are likely to have smaller memory footprints,
resulting in lower overhead should they be materialized to the swap
partition.

\parag{\res Locality} When a suspended job is resumed, we
take care to avoid redundant work by not restarting from scratch tasks
in a suspended state. \htt{}s with free slots that host suspended
tasks for a resumed job will \res those tasks (oldest tasks get
resumed first, because of the reasons just outlined for
suspension). On the other hand, \htt{}s with free slots but no
suspended tasks will only schedule new tasks that have not been
previously scheduled. In other words, suspended tasks will only be
rescheduled as soon as a slot frees up in the \htt they are suspended
on.



\parag{Maximum Number of Suspended Jobs} Suspending tasks
has a cost in terms of storage space requirements. If many tasks on a
single machine are suspended, process data could use a large fraction
of the RAM available on a machine and eventually could also deplete
the swap space. This is an extreme case that only arises with
particular workloads (several jobs arriving in decreasing size); we
address it by defining a threshold on the number of tasks that can be
suspended. When too many tasks are suspended, HFSP switches to the
\wait{}-based preemption technique, until conditions are met for
reverting to eager preemption.

Setting a limit on the number of suspended processes effectively
implies limiting also the total amount of memory allocated to
suspended processes, since in Hadoop the amount of memory allocated
per \reduce task cannot exceed a configurable threshold (the default
is 1 GB)~\cite{cluster_setup}.

\parag{Side effects} Eager preemption should be used with care in case of MapReduce jobs that operate on ``external'' resources, e.g. that heavily use Hadoop \textit{streaming} or \textit{pipes}. Our implementation can be easily extended to provide API support to inhibit \sus{} and \res{} primitives for such particular workloads.

\section{Experiments}
\label{sec:experiments}
This Section focuses on a comparative analysis between FAIR and HFSP schedulers. For the sake of readability -- due to extremely large sojourn times -- we omit the default Hadoop FIFO scheduler from our figures.

Next, we specify the experimental setup and present a series of results, organized in macro and micro benchmarks. Macro benchmarks illustrate the global performance of the schedulers we study in this work, in terms of job sojourn times. Micro benchmarks, instead, focus on the peculiarities of HFSP.

\begin{table*}[t!]
\centering
\begin{tabular}{|r|r|r|r|r|}
  \hline
  {} & {\bf \% of Jobs} & {\bf \# Maps} & {\bf \# Reduces} & {\bf Label}\\
  \hline
  \hline
  {\texttt{FB09}} & { (Small jobs) 53} & {$\leq 2$}         & {0}       & {select} \\ 
  {} & { (Medium jobs) 41} & {$2 < x \leq 500$}         & {$\leq 500$} & {aggregate} \\ 
  {} & { (Big jobs)\ \ 6}  & {$\geq 500$}         & {$\geq 500$} & {transform} \\ 
  \hline
  {\texttt{FB10}} & {39} & {$\leq 1500$} & {$\leq 10$}         & {expand} \\
  {}              & {16} & {$\leq 1500$} & {$10 < x \leq 100$} & {expand and transform}\\
  {}              & {11}  & {$\leq 1500$} & {$x > 100$} & {transform}\\

  {}              & {10} & {$1500 < x \leq 2500$} & {$x \leq 100$} & {aggregate}\\
  {}              & {7} & {$1500 < x \leq 2500$} & {$x > 100$} & {transform}\\

  {}              & {10} & {$x > 2500$} & {$x > 100$} & {transform} \\
  \hline
  \hline
\end{tabular}
\label{tab:workloads}
\caption{Job types in each workload, as generated by SWIM. Jobs are labeled according to the analysis tasks they perform. For the \texttt{FB2009} dataset, jobs are clustered in bins and labeled according to their size.}
\end{table*}

\newpage
\subsection{Experimental Setup}
\label{sec:setup}

In this work we use both a cluster deployed on Amazon EC2 \cite{aws-ec2} -- which we label the Amazon Cluster --  and the standard Hadoop emulator, Mumak \cite{mumak}. The Amazon Cluster is configured as follows: we deploy 100 ``\texttt{m1.xlarge}'' EC2 instances, each with four 2 GHz cores (eight virtual cores), 4 disks that provide roughly 1.6 TB of space, and 15 GB of RAM.\footnote{This is the configuration used by Zaharia \textit{et. al.}~\cite{eurosys10}.} In our experiments -- with the Amazon Cluster and with Mumak -- the HDFS block size is set to 128 MB and a replication factor of 3, while the main Hadoop configuration parameters are as follows: we set 4 \map{} slots and 2 \reduce{} slots per node. 

\parag{Workloads} Generating realistic MapReduce workloads is a difficult task, that has only recently received some attention. In this work, we use SWIM \cite{swim_tool}, a standard benchmarking suite that comprises workload and data generation tools, as described in the literature \cite{swim, workloads_interview, workloads}. A workload expresses in a concise manner \textit{i)} job inter-arrival times, \textit{ii)} a number of \map{} and \reduce{} tasks per job, and \textit{iii)} job characteristics, including the ratio between output and input data for \map{} tasks. 
For our experiments, we use two workloads synthesized from traces collected at Facebook and publicly available on \cite{swim_tool} -- that we label \texttt{FB2009} and \texttt{FB2010} -- as done in previous works \cite{eurosys10, swim}. Table~\ref{tab:workloads} describes the workloads we use, as generated by SWIM for a cluster of 100 nodes.

The \texttt{FB2009} workload comprises 100 unique jobs, and is dominated by small jobs, i.e., jobs that have a small number of \map{} tasks and no \reduce{} tasks. The job inter-arrival time is a random variable with an exponential distribution and a mean of 13 seconds, making the total submission schedule 22 minutes long. Experiments with the \texttt{FB2009} dataset illustrate scheduling performance when the system is not heavily loaded, but has to cope with the demand for interactivity (small jobs).

The \texttt{FB2010} workload comprises 93 unique jobs, which have been selected by instructing SWIM to filter out small jobs. In particular, the number of \map{} tasks is substantially larger than the number of available slots in the system: \map{} phases require multiple ``waves'' to complete. The number of \reduce{} tasks varies between 10 and the number of available reduce slots in the cluster. The job inter-arrival time is a random variable with an exponential distribution and a mean of 38 seconds, making the total submission schedule roughly 1 hour long. The \texttt{FB2010} dataset is particularly demanding in terms of resources: as such, scheduling decisions play a crucial role in terms of system response times.

\parag{Scheduler Configuration} Unless otherwise stated, HFSP operates with the delay scheduling technique and eager preemption enabled. HFSP requires a handful of parameters to be configured, which mainly govern the estimator component: we use a uniform distribution to perform the fitting of the estimate job size;
the sample set size $s$ for \map{} and \reduce{} tasks is set to 5; the parameter $\Delta$ to estimate \reduce{} task size, is set to 60 seconds; we set the parameter $\xi=1$. For the workloads we use in our experiments, the parameters described above give the best results. 

The FAIR scheduler has been configured with a single job pool, using
the parameters suggested in the Hadoop scheduler documentation
\cite{fair_site, slowstart_jira}.

\begin{figure*}[t!]
  \centering
  \begin{tabular}{ccc}
    \begin{minipage}[t]{0.3\textwidth}
      \begin{center}    
        \subfigure[Small jobs]{
          \label{fig:bin1}
          \includegraphics[scale=0.9]{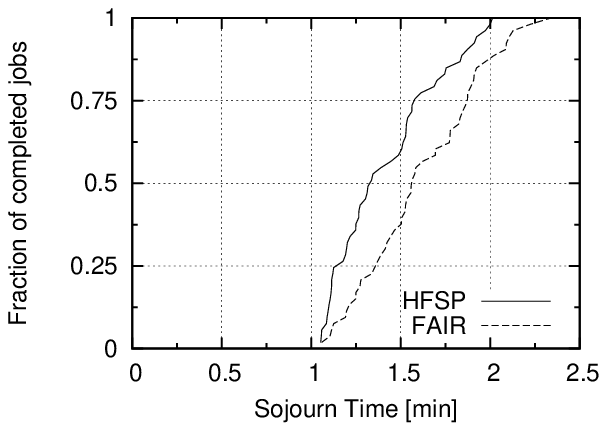}}
        
      \end{center}
    \end{minipage}
    
    &
    
    \begin{minipage}[t]{0.3\textwidth}
      \begin{center}    
        \subfigure[Medium Jobs]{
          \label{fig:bin2}
          \includegraphics[scale=0.9]{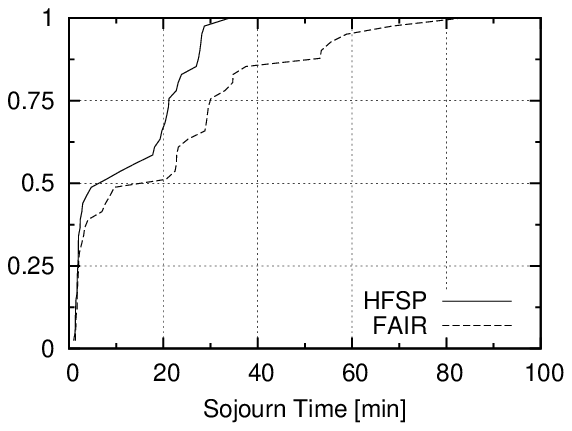}}
      \end{center}
    \end{minipage}

  &

    \begin{minipage}[t]{0.3\textwidth}
      \begin{center}    
        \subfigure[Large Jobs]{
          \label{fig:bin8}
          \includegraphics[scale=0.9]{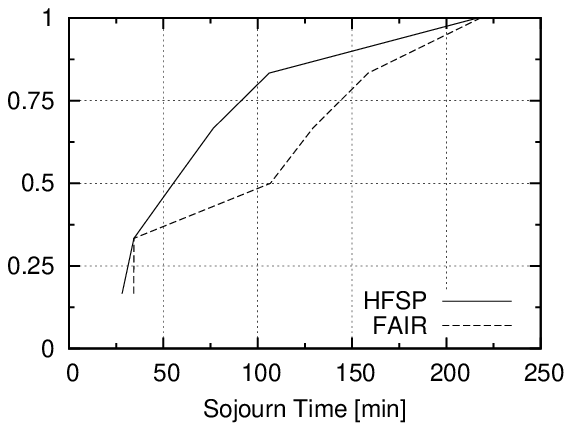}}
      \end{center}
    \end{minipage}
    
  \end{tabular}
  \caption{ECDFs of sojourn times for the \texttt{FB2009} dataset. Jobs are clustered in classes, based on their sizes. HFSP improves the sojourn times in most cases. In particular, for small jobs, HFSP is slightly better than FAIR, whereas for larger jobs, sojourn times are significantly shorter for HFSP than for FAIR. }
  \label{fig:cdf_FB2009}
\end{figure*}

\begin{figure*}[t!]
  \centering
  \begin{tabular}{ccc}
    \begin{minipage}[t]{0.3\textwidth}
      \begin{center}    
        \subfigure[\map{} phase]{
          \label{fig:cdf_map}
          \includegraphics[scale=0.9]{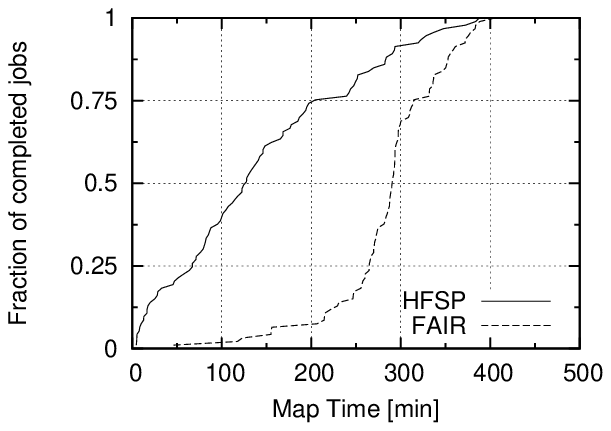}}
        
      \end{center}
    \end{minipage}
    
    &
    
    \begin{minipage}[t]{0.3\textwidth}
      \begin{center}    
        \subfigure[\reduce{} phase]{
          \label{fig:cdf_red}
          \includegraphics[scale=0.9]{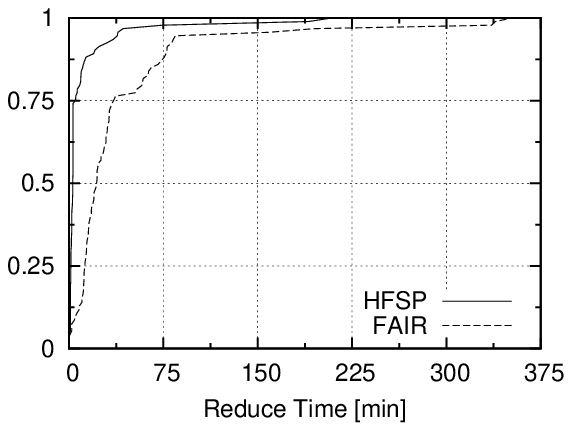}}
      \end{center}
    \end{minipage}

  &

    \begin{minipage}[t]{0.3\textwidth}
      \begin{center}    
        \subfigure[Aggregate]{
          \label{fig:cdf_agg}
          \includegraphics[scale=0.9]{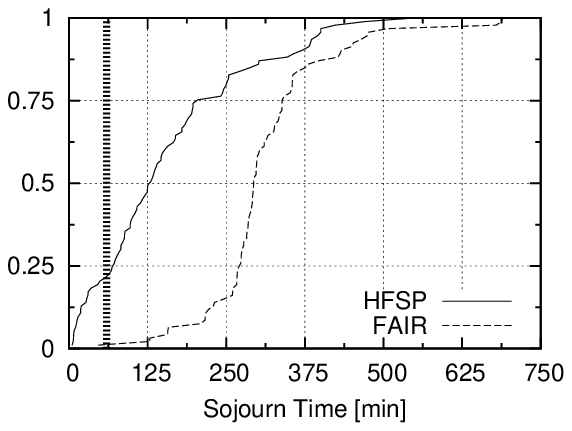}}
      \end{center}
    \end{minipage}
    
  \end{tabular}
  \caption{ECDFs of sojourn times for the \texttt{FB2010} dataset. The
    \map{} phase (left) shows significant improvements of HFSP over
    FAIR. Shorter sojourn times in the \map{} phase are reflected in
    the \reduce{} phase (middle), which shows that HFSP outperforms
    FAIR in terms of sojourn times. The ECDF of aggregate job sojourn
    times (right) indicates that both median and worst-case sojourn
    times are better with HFSP. The vertical line at 60 minutes (the
    workload has been consumed) indicates 20\% of completed jobs with
    HFSP vs. 1\% of completed jobs with FAIR.}
  \label{fig:cdf_FB2010}
\end{figure*}

\subsection{Macro Benchmarks}\label{sec:results}

We now use the Amazon Cluster and report the empirical cumulative distribution function (CDF) of sojourn times for FAIR and HFSP, when the cluster executes the \texttt{FB2009} and \texttt{FB2010} workloads. Although we do not include results we obtain with the FIFO scheduler, for reference, our experiments report a mean sojourn time 5 to 10 times bigger than that of HFSP, depending on the workload.

Figure \ref{fig:cdf_FB2009} groups results according to job sizes: the \texttt{FB2009} dataset contains three distinct clusters of job sizes (small, medium and large), with small jobs dominating the workload. Our results indicate a general improvement of job sojourn times in favor of HFSP. For small jobs, the fair share of cluster resources allocated by both HFSP and FAIR is greater than their requirements in terms of number of tasks. In addition, very small jobs (with 1-2 \map{} tasks only) are scheduled as soon as a slot becomes free (both under the HFSP and FAIR), and therefore their sojourn time depends mainly on the frequency at which slots free-up and on the cluster state upon job arrival. Overall, we observe that HFSP performs slightly better than FAIR for small jobs. For medium and large jobs, instead, an individual job may require a significant amount of cluster resources. Thus, the advantage of HFSP is mainly due to its ability to ``focus'' cluster resources -- as opposed to ``splitting'' them according to FAIR -- towards the completion of the smallest job waiting to be served. 

Figure \ref{fig:cdf_FB2010} shows the results for the \texttt{FB2010} dataset. We show the sojourn times of the \map{} and \reduce{} phases, and the total job sojourn times. In the \map{} phase (Figure~\ref{fig:cdf_map}), the sojourn times are smaller for HFSP than for FAIR: the median sojourn time is more than halved. This is a consequence of the HFSP discipline, which dedicates cluster resources to individual jobs rather than spreading them on multiple ``waves''. 
As such, with HFSP, \reduce{} tasks enter the \shuffle{} phase earlier than what happens with FAIR, and have to wait less time for all their input data to be available. Therefore, the \reduce{} phase is shorter with HFSP, as shown in Figure~\ref{fig:cdf_red}. Clearly, the HFSP discipline in the \reduce{} phase also contributes to short sojourn times, with a median difference of roughly 30 minutes. Finally, the total job sojourn times indicate that the system response time with HFSP substantially improves. For illustrative purposes, we show a vertical line corresponding to 60 minutes, by which all jobs of the \texttt{FB2010} workload arrived in the system. By that time, only one job completes with FAIR, whereas more than 20\% of the jobs, are served with HFSP. More to the point, when 80\% of jobs are served by HFSP, roughly 15\% of jobs are served with FAIR.

In summary, HFSP caters both to workloads geared towards ``interactive'' small jobs and to a more efficient allocation of cluster resources, which is beneficial to large jobs. Next, we focus on HFSP in particular and study the behavior of its inner components in detail.

\subsection{Micro Benchmarks}\label{sec:micro}

In the following set of experiments, when not otherwise stated, we use the Mumak emulator to execute the workloads generated by SWIM and described earlier; in addition, we use synthetic workloads that reproduce peculiar cases to better explain how HFSP works.

\parag{Impact of Cluster Size} The goal of this experiment is to study the behavior of both HFSP and FAIR, when cluster resources available for scheduling decisions are scarce. Indeed, many real-life deployments \cite{hadoop_power_by} are smaller than 100 nodes, yet they are used to execute a large number of jobs on big data volumes. We use the \texttt{FB2009} dataset, because it contains a mix of different job sizes, including small jobs.
In our experiments, we vary the cluster size in the range of 10 to 100 machines. When the cluster size is small, we increase accordingly the storage space available at each node, to accommodate the data volume used in the \texttt{FB2009} workload. 

\begin{figure}[htbp]
  \centering
  \includegraphics[scale=1]{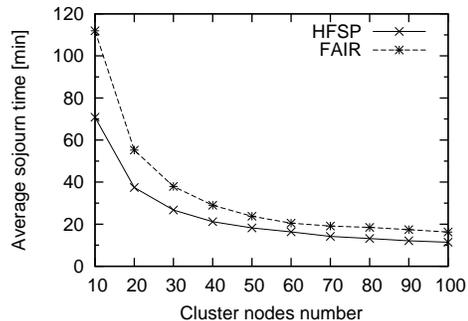}
  \caption{Impact of cluster size on scheduling performance, for both HFSP and FAIR, using the \texttt{FB2009} dataset.}
  \label{fig:cluster_size}
\end{figure}

Figure~\ref{fig:cluster_size} reports the mean job sojourn time for HFSP and FAIR, as a function of the number of machines in the cluster. When resources are scarce, HFSP achieves considerably smaller mean job sojourn times due to its ability to ``focus'' cluster resources to individual jobs. Put otherwise, for equivalent job sojourn times, the \texttt{FB2009} workload requires a smaller cluster when HFSP is used, as compared to other schedulers, which is a desirable property as it may relate to significant cost savings. Clearly, our results hold (and the gap in favor of HFSP is even more substantial), when the \texttt{FB2010} dataset is used.

\parag{Robustness to Estimation Errors} In this experiments we study whether HFSP incurs performance degradation when job size -- an essential ingredient for the correct operation of the scheduler -- is incorrect. We do so by injecting artificial errors on the job size estimates reported by the Training module of HFSP, and measure the impact in terms of mean job sojourn times. 

For this experiment, we focus only on errors injected in the estimator of the \map{} phase size. To do so, we use a modified, \map{} only version of the \texttt{FB2009} dataset. In practice, the \map{} size estimate used by the virtual cluster of HFSP is a random variable uniformly distributed in the range $[\theta() \cdot (1-\alpha), \theta() \cdot (1+\alpha)]$, where $\theta()$ is the size estimate computed as in Section~\ref{sec:training}, and $\alpha \in [0.1,1]$ is the artificial error we inject. We repeat each experiment 20 times for each value of $\alpha$, to gain statistical confidence in our results. 

\begin{figure}[htbp]
  \centering
  \includegraphics[scale=1]{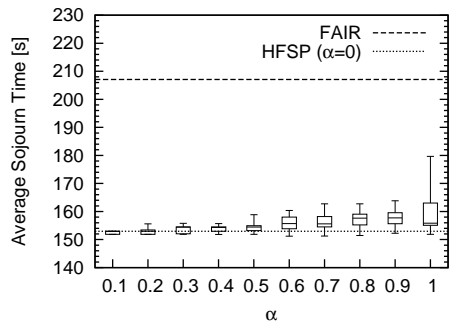}
  \caption{Impact of job size estimation errors on HFSP performance, using a \map{} only version of the \texttt{FB2009} dataset.}
  \label{fig:job-size-errors}
\end{figure}

Figure~\ref{fig:job-size-errors} reports the mean sojourn time for
different $\alpha$ values. In addition, we show as a reference the
mean sojourn time for the same experiments using FAIR, which are
clearly not affected by estimation errors, and the mean sojourn time
achieved by HFSP with $\alpha = 0$. In our experiments, HFSP is
particularly resilient to wrong estimates: indeed, the mean sojourn
times is slightly affected only for extremely large
errors. Indeed, macroscopic changes in sojourn time happen
  only when long jobs are scheduled before much shorter ones, which
  would then incur a large sojourn time. In the \texttt{FB2009}
  dataset -- which is grouped in classes according to job size --
  ``reversals'' (i.e., jobs that are scheduled in a different order)
  only happen within the same class, with only a modest impact on
  sojourn times.


\parag{Data Locality} Next, we focus on data locality and measure the
fraction of tasks that read data from the local disk of the machine
they run on.

We use the same experiments described in Section~\ref{sec:results}, where we use the Amazon Cluster to execute the \texttt{FB2009} and \texttt{FB2010} datasets, for both HFSP and FAIR. For the sake of brevity, we only report results for the \texttt{FB2009} dataset.

FAIR achieves 98\% of data locality whereas HFSP always achieves 100\% of data locality, over a total of more than 14,000 tasks across all experiments. Clearly, the delay scheduler mechanism \cite{eurosys10} is beneficial to both FAIR and HFSP. Additionally, we observe that the result we obtain is also a consequence of resource allocation: with HFSP, a job scheduled for execution receives (if the cluster size allows it) all the resources required for its processing, whereas with FAIR, it is granted fewer resources. As a consequence HFSP copes better with the random data placement strategy used by HDFS, and obtains more local tasks, which contributes to shorter job execution times and hence smaller sojourn times. 

\subsection{Job Preemption}
\label{sec:preemption_study} 

We now study in detail the various preemption mechanisms we discuss in Section~\ref{sec:preemption}. First, we compare the behavior of HFSP with eager preemption, with the \wait{} primitive, and with a \kil{} primitive, for a simple workload. The goal of such experiment is to illustrate the benefits of eager preemption, in terms of job sojourn times, and to discuss when alternative mechanisms are to be preferred. Then, we move to another set of experiments, in which we study the performance of the \sus{} and \res{} operations, executed in a small local cluster. 

For the first set of experiments we use a simple, synthetic workload composed of five jobs, and focus solely on \reduce{} tasks, since \map{} tasks are not preempted with HFSP. We set up a small cluster in Mumak, with 4 machines with 2 \reduce{} slots each. In our workload, the first job, $j_1$, has 11 reduce tasks each of duration roughly 500 seconds and arrives at time 2 minutes and 20 seconds. All the other jobs arrive at time 2 minutes and 30 seconds and all have one \reduce{} task, except for $j_2$ that has two \reduce{} tasks. For jobs $j_2 \cdots j_5$, \reduce{} task times are smaller than the one of $j_1$.

\begin{figure}[t]
  \begin{tabular}{c}
    \begin{minipage}[t]{0.29\textwidth}
      \begin{center}    
        \subfigure[HFSP with eager preemption]{
          \label{fig:susp}
        \includegraphics[scale=.23]{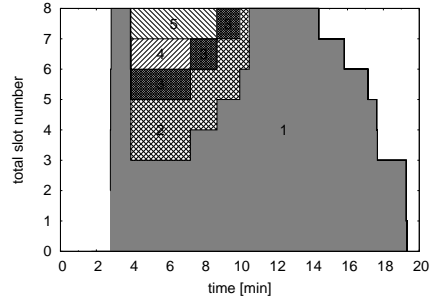}}
      \end{center}
    \end{minipage}
    
    \\
    
    \begin{minipage}[t]{0.29\textwidth}
      \begin{center}    
        \subfigure[HFSP with \wait{}]{
          \label{fig:nosusp}
        \includegraphics[scale=.23]{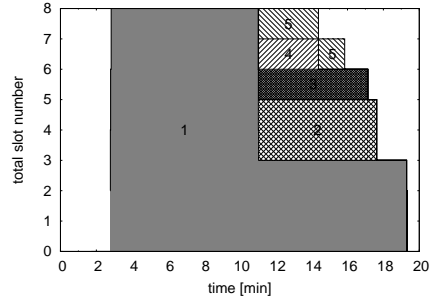}}
      \end{center}
    \end{minipage}

  \end{tabular}
  \caption{Resource allocation graphs for a simple workload, with and without eager preemption.}
  \label{fig:susp2}
\end{figure}

Figure~\ref{fig:susp2} illustrates a resource allocation graph that we obtain by processing Hadoop log files: on the y-axis we report the cumulative slot utilization per job, on the x-axis we report time, in minutes. In Figure~\ref{fig:susp}, which shows the behavior of HFSP with eager preemption, when jobs $j_2,j_3,j_4$ and $j_4$ arrive, they preempt $j_1$ and occupy the cluster with their tasks. Note that HFSP suspends only the required number of tasks of $j_1$ to accommodate the newly arrived jobs. When jobs $j_2 \cdots j_5$, the suspended tasks of job $j_1$ are resumed. The average sojourn time in this simple example is about 9 minutes. Instead, in Figure.~\ref{fig:nosusp}, in which HFSP uses the \wait{} primitive, when jobs $j_2,j_3,j_4$ and $j_4$ arrive, the cluster is fully occupied by $j_1$. As such, HFSP waits for job $j_1$ to complete the required number of tasks necessary to allocate the new jobs, before proceeding with scheduling. As a consequence, the average sojourn time is 15 minutes, roughly 40\% larger than with preemption. We also repeat the very same experiment by implementing a simple \kil{} primitive: in this case, job $j_1$ has a larger completion time because 6 of its tasks are killed due to the arrival of jobs $j_2 \cdots j_5$, and thus need to be scheduled again. We omit the resource allocation graph for this case, as it is very similar to that in Figure~\ref{fig:susp}.

It is possible to define alternative scenarios in which HFSP could
achieve better sojourn times with the \wait{} primitive. In general,
when task runtimes are short, the \wait{} primitive is to be
preferred: eager preemption may need to perform process swapping,
which could take longer than the remaining time for the task to
complete. Instead, when task runtimes are long, eager preemption is a
more sensible choice, as it brings shorter sojourn times.

Next, we study the time it takes, for the OS, to perform the \sus{} and \res{} operations, as this determines to a great extent whether eager preemption is to be preferred over the \wait{} primitive.


\parag{Swap Times} Recall that, in our implementation of HFSP, when one or more tasks of a job are preempted, the memory that they are using can be claimed by other tasks scheduled to occupy their slot. In this case, the Operating System (OS) may swap the memory contents to disk. When such preempted task are resumed, the OS reloads in memory the swapped process from disk. As such, the \sus{} and \res{} operations may introduce delays that we seek to compare to task times.
We remark that such delays are bounded: indeed, the memory footprint of a task is limited by the way a MapReduce job is engineered. When a task is preempted, the amount of memory it uses is bounded by the amount of RAM per slot, a parameter configured in Hadoop. As such, the disk I/O that characterizes cluster machines, together with the amount of RAM used by \reduce{} tasks, are the main factors to consider when configuring HFSP to use eager preemption.

To verify our claim, we perform the following experiment on a local cluster, with a single \htt{} and a single \reduce{} slot. The hardware configuration we use is as follows: the \htt{} machine has 8 GB of RAM, and a local 7.2Krpm disk with a read/write I/O speed of roughly 100 MBps. The Linux ``swappiness'' \cite{swappiness} is configured to a small (5) value; in addition, swap size is configured such that it can accommodate several memory dumps.

We define a simple workload\footnote{We omit \map{} tasks, as they are not preempted with HFSP.} in which a job with a single \reduce{} task occupies 6 GB of RAM (with randomly generated data), and the heap space of the child JVM running the task is set to 6 GB. Then, a new job with a single \reduce{} task arrives in the system, and preempts the initial job. The ``swap'' time is defined as the time it takes for the system to \sus{} the initial \reduce{} task and to \res{} it. The new job is characterized by a variable memory footprint, varying from 0.5 GB up to 6 GB.

\begin{figure}[htbp]
  \centering
  \includegraphics[scale=1]{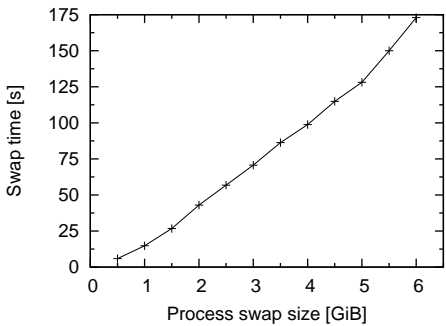}
  \caption{Swap times for a small cluster, using a synthetic workload of two jobs with one \reduce{} task each. The second \reduce{} task has a variable memory footprint, which corresponds to the amount of data swapped to disk by the OS, as reported on the x-axis.}
  \label{fig:swap}
\end{figure}

Figure~\ref{fig:swap} reports the swap time for the experimental setup described above. Swap time is approximately linear in the size of the memory footprint of the preempting task, up to a process swap size of 5 GB. Our results indicate that the \sus{} and \res{} primitives operates roughly at disk raw-speed: for example, for a process swap size of 2 GB, the systems writes, and then reads 2 GB in roughly 40 seconds, giving an I/O speed of roughly 100 MBps. Then, swap performance slightly decreases after 5 GB, which accounts for the overhead of the OS trying to balance resources dedicated to the swap partition and to caching.

Putting the results in perspective, we now refer to the experiments with the \texttt{FB2010} dataset described in Section~\ref{sec:results}. Our logs indicate that at most 10\% of running tasks have been suspended and then resumed: such tasks belong to 10 different jobs of the workload. In those experiments, the JVM heap space for a \reduce{} task is set to 1 GB \cite{cluster_setup}, which means that \sus{} and \res{} operations induce at most a swap time around 20 s. Since the average duration of preempted tasks is one order of magnitude larger than the swap time, we conclude that eager preemption is a sensible configuration for HFSP with the workload we used.

\section{Discussion}
\label{sec:discussion}

We now discuss several points that complement the work we have presented so far.

\parag{Preemption Performance} Some works have considered job preemption primitives alternative to the ones we propose in this work. For example, Cheng \textit{et al.}~\cite{killmr} present a detailed analysis of a \kil{} primitive to implement job preemption and come up with a method to select the best tasks to kill to avoid hurting too much job performance. Instead, Anantharanayanan \textit{et al.}~\cite{socc12} observe that job preemption can be implemented at the ``application-level'', meaning that individual tasks may be suspended and resumed by Hadoop, which would require a dedicated memory-management module to handle preemption.

One reason to avoid using OS paging might be that it could affect system stability \cite{socc12}: a large number of interactive applications that require more RAM than that available on the system, substantially hinder the task of the OS. 

Given the results we show in Section~\ref{sec:preemption_study}, it is reasonable to assume that an ``application-level'' preemption mechanism cannot perform better in moving data from/to RAM and disk, than what we propose in our work. In addition, Hadoop jobs are not interactive applications that would need frequent context switching. As such, while we do not dismiss any effort to study alternative mechanisms, we believe the eager preemption we present in this work to be effective and practical in achieving its goals.


\parag{Job with Different Priorities}  The design of HFSP takes as a reference the Processor Sharing (PS) discipline to compute the order of the jobs to be scheduled. In PS, each job receives its equal share of the resources. A natural extension of the work would provide different priorities, or weights, to jobs: in this case, we shall consider the Generalized Processor Sharing (GPS) discipline, where each job receives an amount of resources in proportion to its weight. For instance, if $\mathcal{J}$ is the set with all the jobs in the system, then job $k$ with weight $w_k$ will receive a fraction $\displaystyle \frac{w_k}{\sum_{i \in \mathcal{J}} w_i} $ of the resources. This computation can be easily incorporated in the job aging computation done by the HFSP algorithm.

\parag{Job Size Estimation} In some cases, task execution time, which contributes to job duration, could be regarded as a highly variable quantity, making task time distributions highly skewed, thus difficult to estimate. 

We remark that in HFSP, the estimator is designed as a pluggable module that can be replaced by more sophisticated estimation techniques, therefore providing more accurate predictions. 
In addition, in recent work Kwon~\textit{et al.}~\cite{skewtune} address and greatly mitigate the issue of skew in task processing times with a plug-in module that seamlessly integrates in Hadoop, which can be used in conjunction with HFSP. Moreover, Popescu~\textit{et al.}~\cite{query_perf} present an appealing approach to predict MapReduce ``query'' runtime, that can be also used in HFSP.
We conclude by remarking that the original FSP discipline has also been studied in the case of inaccurate job size information \cite{inaccuratesizebasedscheduling}: according to such work, FSP is a stable algorithm that is robust to inaccurate job size, a result that we confirm in the context of this paper.

\section{Related Work}
\label{sec:related_work}

MapReduce in general and Hadoop in particular have received a lot of attention recently, both from the industry and from academia. In this work we focus on job scheduling, and consider the literature pertaining this domain. 
Several theoretical works tackle scheduling problems in a multi-processor system -- see for instance \cite{moseley_soda}. These works, which represent elegant and important contributions to the domain, consider jobs with a simple structure (\textit{i.e.,} a single phase) and make several simplifying assumptions on the underlying execution system. The main objective of such theoretical studies is to offer bounds on job performance provide optimality results. In contrast, in this work we take a system approach, and focus on the design and implementation of a scheduling mechanism taking into account all the details and intricacies of a real system.

More recently, the problem of job scheduling in MapReduce has revived interest in theoretical approaches to study job performance. Some works~\cite{infocom11, moseley_spaa} provide interesting approximability results but fail in providing a truthful model of the underlying MapReduce system. In the same vein, but with results that are readily applicable, Tan \textit{et al.}~\cite{sigm12} identify several shortcomings of the FAIR scheduler we also study in this work and proposes an elegant model of job runtimes. Their contribution aims at mitigating job starvation problems that arise when job runtimes are heavily skewed. In contrast, our goal is, more generally, to overcome problems of processor-sharing disciplines with respect to job sojourn times. As such, the results of Tan \textit{et al.} could be extended to cover our scheduler.

The works that are more closely related to ours, because they have a system approach to scheduling and aim at the design and implementation of a scheduling discipline, are numerous. For instance, the FAIR scheduler and its enhancement with a delay scheduler \cite{eurosys10} is a prominent example to which we compare our results. Another work by Tan~\textit{et al.}~\cite{infocom12} provides more system details on the mechanism used to overcome job starvation with the FAIR scheduler. Several other works \cite{sigm09, quincy09, nsdi11, nsdi11b} focus on resource allocation and strive at achieving fairness across jobs, but do not consider sojourn times. Sandholm and Lai~\cite{dynamic} study the resource assignment problem through the lenses of a bidding system to achieve a dynamic priority system and implement quality of service for jobs. Kc and Anyanwu~\cite{deadlines} address the problem of scheduling jobs to meet user-provided deadlines, but assume job runtime to be an input to the scheduler. Finally, the work that is more closely related to ours is Flex~\cite{flex}, which provides a framework for the optimization of any given performance metric. In particular, when the performance metric is chosen to be the ``max-sum'' sojourn-time, Flex should minimize the average sojourn time, whereas in our work we cannot make any optimality claims. Flex is implemented as an add-on on top of the FAIR scheduler, and shares similar design principles to our work. For example, when configured to operate as a size-based scheduler, Flex implements an estimation module to infer job sizes.
Unfortunately, we could not compare our proposal to Flex because it is a proprietary software without a sufficient documentation describing its internals.

Finally, various recent approaches~\cite{ARIA11, mascots12, nsdi12-c,
  query_perf, mrpredict} have been proposed to infer the size of
MapReduce jobs for specific application scenarios.
HFSP is designed such that the estimator module can be easily plugged
with other mechanisms, benefiting from advanced and tailored solutions.

\section{Conclusion}
\label{sec:conclusion}

The problem of scheduling jobs in parallel systems have received a lot of attention in the past, including works that attempted at producing elegant mathematical models of such systems with the goal of studying the hardness of obtaining optimal scheduling. In this work we took a systems approach, glossing over mathematical constructs and optimality analysis: instead we were interested in studying the benefits of a size-based approach to scheduling jobs in a real system, namely Hadoop.

Our work was motivated by the realization that MapReduce has evolved to the point where shared clusters are used for a wide range of workloads, which include an increasingly large fraction of interactive data processing tasks. Existing schedulers in the state-of-the-art suggest, to overcome the inherent limitations of a simple first-come-first-served discipline, cluster resources to be shared equally among running jobs. As a consequence, we have witnessed the raise of deployment best practices in which long sojourn times were compensated by over-dimensioned Hadoop clusters. Armed with the realization that a large fraction of cluster resources were used for a small amount of time, given a selection of real-world workload traces, in this work we set off to study the benefits of a new scheduling discipline that targeted at the same time short sojourn times and fairness among jobs.

The HFSP scheduler we proposed in this article brought up several challenges. First, we came up with a general architecture to realize \textit{practical} size-based scheduling disciplines, where job size is not assumed to be known a priori. The HFSP scheduling algorithm solved many problems related to the underlying discrete nature of cluster resources, how to keep track of jobs making progress towards their completion, and how to implement eager preemption primitives. Then, we used statistical tools to infer task time distributions and came up with an approach aiming at avoiding wasting cluster resources while estimating job sizes. Finally, we performed a comparative analysis of HFSP with the two standard schedulers that are most widely used today in production-level Hadoop deployments, and showed that HFSP brings several benefits in terms of shorter sojourn-times, even in small, highly utilized clusters.

There are several avenues that we are considering as part of our future work. First, we will extend our experimental study to cover a wider range of workloads, including those presenting issues related to skew in task time distributions. In this work, on of our goal was to avoid wasting work done, both during the training phase, and by using eager preemption. We plan to investigate how to integrate our preemption mechanism with the ``speculative execution'' mode of Hadoop, since in some cases, it could be beneficial to sacrifice some work done by a job to gain in system reactivity for scheduling decisions.
Finally, we will study the problem of scheduling complex job workflows, that result from the composition of several sub-jobs. 




\balance
\bibliographystyle{abbrv}
\bibliography{references}

\end{document}